\documentclass[twocolumn,preprintnumbers,amsmath,amssymb,aps,prb]{revtex4}
\usepackage{graphicx}
\begin{document}

\title{
Active Microrheology in Active Matter Systems: 
Mobility, Intermittency and Avalanches     
} 
\author{
C. Reichhardt and  C. J. Olson Reichhardt}
\affiliation{
Theoretical Division,
Los Alamos National Laboratory, Los Alamos, New Mexico 87545 USA
} 

\date{\today}
\begin{abstract}
We examine the mobility and velocity fluctuations of a driven particle moving 
through an active matter bath 
of self-mobile disks for varied density or
area coverage and varied activity.
We show that the driven particle mobility 
can exhibit non-monotonic 
behavior that is correlated with distinct changes 
in the spatio-temporal
structures that arise in the active media. 
We demonstrate that the probe particle velocity distributions 
exhibit specific features in the different dynamic regimes, and 
identify an activity-induced uniform crystallization that occurs for moderate
activity levels and that is
distinct from the previously observed higher activity cluster phase.  
The velocity distribution
in the cluster phase
has telegraph noise characteristics produced when
the probe particle moves alternately through
high mobility areas that are in the gas state
and
low mobility areas that are in the dense phase.   
For higher 
densities
and large activities, 
the system enters what we characterize as an active jamming regime.
Here
the probe particle moves in intermittent jumps or avalanches which have 
power-law distributed sizes that are
similar to the avalanche distributions
observed for non-active disk systems near the jamming transition.
\end{abstract}
\maketitle

\vskip2pc

\section{Introduction}

In active microrheology, the properties of a medium are probed using
the mobility and velocity fluctuations
of an externally driven probe particle that
is roughly the same
size as the particles that comprise the medium \cite{1,2,3,4}.  
For example, the nonlinear mobility of a probe particle dragged through a 
colloidal system changes across the glass transition
\cite{5,6,7,8}.
Other studies have examined 
anomalous diffusion properties 
of the probe \cite{9,10}, 
features of the driven particle velocity fluctuations, 
and 
threshold-to-motion or depinning type phenomena \cite{11,12,13}.  
In ordered systems, a driven particle can 
induce localized melting \cite{14} or shear
thinning effects \cite{15}.       
Driven particles or intruders have also been used 
to study the onset of jamming as 
a function of the system density\cite{16,17,18,19,20}.
As the jamming transition is approached,
the particle mobility is strongly reduced,
its motion becomes increasingly intermittent, and
just at jamming the velocity fluctuations 
become power-law distributed \cite{16,17,18}.

Most active microrheology studies
have focused on systems where the particles comprising the medium
are not driven or experience only thermal fluctuations.
There is, however, another class of systems to which
the techniques of active microrheology can be applied: 
collections
of self-motile particles or active matter \cite{21a,21b}. 
Examples of such systems include 
run-and-tumble bacteria \cite{22,23a,23b,23c} or
self-mobile colloidal particles \cite{24a,24b,24c,24d,25,26,27}, which are
commonly modeled as 
self-mobile sterically interacting particles undergoing
either active Brownian motion or 
run-and-tumble dynamics.  
Such systems exhibit
a transition from a uniform liquid state at low 
activities and 
densities to a cluster or phase separated state
at higher activities and densities,
where close-packed clusters are surrounded by a low density gas 
\cite{26,27,28,29,30,31,32a,32b,33}. 
For monodisperse particles confined to two dimensions,
the clusters have local triangular ordering, 
and the resulting crystallites can move, break apart, and re-form
\cite{26,27,31,32a,32b,33}. 
A driven probe particle in an active matter system 
should show clear changes in mobility or velocity fluctuations depending on 
the spatio-temporal behavior exhibited by the active matter,
and thus could serve as a powerful tool for understanding 
a wide range of active systems. 
In numerical and theoretical studies 
of a driven probe particle in
an active nematic system, anomalous viscosity effects such
as a negative drag were predicted \cite{34}. 
Experiments on cargo transport through crowded living cells 
showed that motion occurs in intermittent bursts 
which exhibit scaling behavior similar to that 
found in critical jammed solids \cite{35}.   
  
Here we examine the dynamics of a probe particle driven
through a bath of run-and-tumble sterically interacting 
particles for varied
activity or run length and 
varied particle density, 
the term we use for the area fraction covered by the disks. 
We show that the probe mobility and velocity fluctuations are
correlated with distinct dynamic spatial structures 
that form in the bath.
At low densities the probe mobility
decreases with increasing run length, 
while the transition from a uniform active liquid to a phase separated  
cluster state coincides with a pronounced mobility drop. 
At high densities
the mobility becomes an increasingly non-monotonic function of the activity.
When the activity is low and the density is high, the system is disordered
and the mobility is high.  When the run length is increased slightly until it
nearly matches the average spacing between the surfaces of neighboring 
particles, a uniform nearly crystalline state emerges that is correlated
with a drop in the mobility by several orders of magnitude.  As the run length
is further increased, this crystalline state becomes disordered, the
system enters a liquid phase, and the mobility increases by an order of
magnitude or more.  For large run lengths the system transitions from
the liquid state to a phase-separated state, in which the mobility drops 
again.
At high density and the largest run lengths, 
the system forms what we call an actively jammed state 
characterized by strongly reduced mobility. 
Here the probe velocity often drops to zero and 
the probe moves only in
intermittent bursts or avalanches with 
size distributions that can
be fit to a power law. 
These results suggest that, at least in the high density regime,
the addition of activity 
can induce critical behavior similar to the type associated with 
jamming in granular systems \cite{16,17,18,37a,37b}. 

\begin{figure}
\includegraphics[width=3.5in]{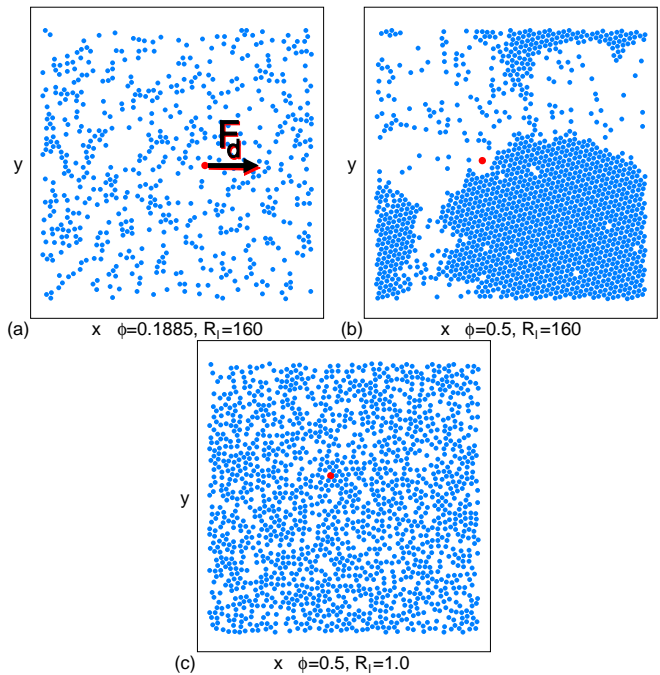}
\caption{ 
Particle locations for an active matter system with
an externally driven probe particle (red).
(a) Uniform liquid state
at $\phi =0.1885$ and $R_{l} = 160$. 
Arrow: direction of the
probe driving force $F_d$.  
(b) Phase-separated cluster state at
$\phi= 0.5$ and $R_{l} = 160$
consisting of a high density phase with local
crystal ordering coexisting with
a low density liquid phase. 
(c) The uniform liquid phase at $\phi = 0.5$ and $R_{l} = 1.0$ 
}
\label{fig:1}
\end{figure}

\section{Simulation} 
We consider a two-dimensional system of size
$L \times L$ with periodic boundary conditions
containing $N$ run-and-tumble monodisperse disks 
of radius $r_d$ that interact via a repulsive harmonic potential. 
The density of particles is given by $\phi=N\pi r_d^2/L^2$,
where we take $L=50$ and $r_d=0.5$ in dimensionless simulation units. 
In the absence of activity, a hexagonal solid forms when $\phi > 0.9$.
The dynamics of a single particle $i$     
is obtained by integrating the overdamped equation of motion 
\begin{equation}
\eta 
\frac{d{\bf R}_i}{dt}
= 
{\bf F}^{m}_{i} +  {\bf F}^{s}_{i} \ .
\end{equation}
Here $\eta = 1.0$ is the damping constant and 
${\bf F}^{m}_{i}$ is the motor force,
which drives 
the particle
in a fixed randomly chosen direction 
under a force of magnitude $F_{m}=0.5$ during a run time $\tau_{r}$. 
A new running direction is randomly selected after each run time.
The run length $R_{l} \equiv F_{m}\tau_{r}$ is 
the distance 
the particle would move in the absence of other particles.
The steric interactions are given by
${\bf F}^{s}_i = \sum^{N}_{i \neq j}(k/r_d)(R_{\rm eff} - |{\bf r}_{ij}|)
\Theta(R_{\rm eff} -|{\bf r}_{ij}|){\hat {\bf r}}_{ij}$,
where ${\bf r}_{ij} = {\bf R}_{i} - {\bf R}_{j}$, 
${\hat {\bf r}}_{ij} = {\bf r}_{ij}/|{\bf r}_{ij}|$, 
$k=20$, $R_{\rm eff} = 2r_{d}$, 
and ${\bf R}_{i(j)}$ is the location of particle $i(j)$.
We add a single non-active probe particle with $F^m=0$ 
to the system with the same radius and 
steric interactions as the active particles and 
apply a constant force ${\bf F}_{d}=F_d{\bf {\hat x}}$ 
with $F_d=0.5$
to only
the probe particle, as shown in Fig.~\ref{fig:1}(a).
Since we are using overdamped dynamics, we employ a forward Euler method
to advance the equations of motion 
with an integration step of $\delta t = 0.001$ 
dimensionless simulation time units.
We have considered other values of $F_{m}$ and 
find that, outside the limits
$F_{d} \gg F_{m}$ or $F_{d} \ll F_{m}$, 
the general features of our results are robust
against the exact choice of $F_d/F_m$.  
We initialize the system by placings the particles in 
non-overlapping randomly chosen positions. 
In active matter particle systems, it is known that in the 
regime where the system forms clumps or becomes phase
separated, there can be long transient times 
before the sample reaches a steady state \cite{31,New}.   
In all cases, prior to taking measurements we wait a sufficient
time for the system to settle into a steady state, as determined
by measuring when the probe particle velocity and the
fraction of sixfold coordinated particles 
$P_6=\sum_i^N\delta(z_i-6)$, where
$z_i$ is the coordination number of an individual particle as obtained from
a Voronoi construction,
settle onto a steady state value.
The length of the
transient time depends on 
the density and run length, and ranges from
$10^3$ simulation time steps in the liquid phase 
up to $10^6$ or more simulation time steps in the dense phase. 
We note that we consider systems containing up to $N=11000$ particles,
which limits the size of the transient times 
compared to the 
considerably larger number of particles employed in \cite{31,New}.  

To characterize the system we measure the time series of the probe velocity fluctuations
$V_x(t)=(d{\bf R}_p/dt)\cdot {\bf \hat x}$,
where ${\bf R}_p$ is the probe position.
In the absence of any other particles,
$F_d/\langle V_x\rangle=1.0$.
In previous simulations of 
the same run-and-tumble system performed with
no probe particle, 
a transition from a liquid state to a cluster state 
occurred for fixed $\phi$ and increasing $R_{l}$, or for
fixed $R_{l}$ and increasing $\phi$ \cite{33}. 
Very similar results 
have been obtained for active Brownian particles,
where at a fixed density a transition to a cluster state 
occurs for increasing persistence length \cite{30,31}.
It has been shown that run-and-tumble dynamics and active Brownian 
motion produce equivalent 
results when the mobility of the particles is density dependent
\cite{36}, which occurs when particle-particle interactions are present,
so we expect our results to be general to either type of system.  

\section {Probe Mobility vs Run lengths and $\phi$ }

We first consider the mobility of the probe particle by fixing
$F_{d}$ and conducting a series of simulations for varied 
particle density $\phi$ and
varied run length $R_{l}$. 
As a point of reference, note that nonactive disks 
with $R_l=0$ form a triangular solid
at a density of $\phi = 0.9$.
In Fig.~\ref{fig:1}(a) we show snapshots of the active 
and drive particle positions 
in a uniform liquid state at $\phi = 0.1885$ and $R_{l} = 160$.
When the run length is held constant but the density is increased,
the system no longer remains a uniform
liquid but instead forms a 
phase separated or cluster state, as shown in 
Fig.~\ref{fig:1}(b) at $\phi=0.5$ and $R_l=160$.
Within an individual cluster, the local density $\phi_{\rm loc}$ is
just below $\phi_{\rm loc}=0.9$ and there is considerable hexagonal ordering
of the particles.  
The clusters are not static but gradually change
over time.
For $\phi  = 0.5$, the cluster state appears when
$R_{l} > 30$. 
Figure~1(c) illustrates a sample with
$\phi = 0.5$ and
$R_{l} = 1.0$. 
Here the run length is small enough that 
the clusters are lost and a uniform liquid reappears.

\begin{figure}
\includegraphics[width=3.5in]{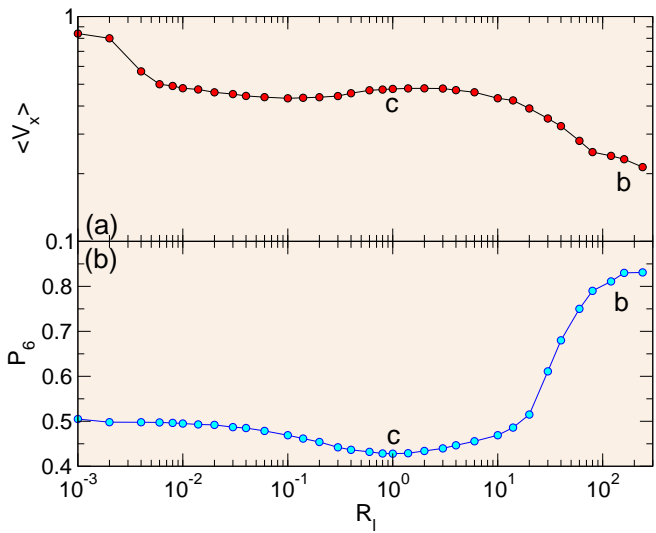}
\caption{ 
(a) Mobility $\langle V_x\rangle$ of the probe particle
vs run length $R_{l}$ for 
the system from Fig.~1(b,c) with $\phi = 0.5$. 
(b) 
Corresponding fraction of six-fold coordinated particles $P_{6}$ vs $R_{l}$.
The $R_l=160$ point illustrated in Fig.~\ref{fig:1}(b), where the system is
phase separated and the probe particle mobility is low, is marked {\bf b},
while the $R_l=1$ point illustrated in Fig.~\ref{fig:1}(c), where the
system is in a liquid state and the mobility is higher, is marked {\bf c}.
}
\label{fig:2}
\end{figure}

In Fig.~\ref{fig:2}(a) we plot the mobility 
$\langle V_x\rangle$ of the probe particle as a function of $R_l$ for
the system 
with $\phi=0.5$ illustrated in Fig.~\ref{fig:1}(b,c).
In order to correlate the mobility with 
different dynamical structures in the active media,  
in Fig.~\ref{fig:2}(b) we plot the corresponding 
fraction of sixfold-coordinated particles 
$P_6$ versus $R_l$.
In the disordered liquid state for low $R_l$, $P_6$ is low,
while in the cluster state $P_6>0.8$ since, within the clusters,
most particles have six neighbors.
The mobility initially decreases with increasing $R_l$ in the range
$0.001 < R_l < 0.01$, passing through
a local minimum near $R_{l} = 0.1$. 
For $ 0.3 < R_{l} < 8$ there is small increase in $\langle V_x\rangle$,
while
for $R_{l} > 8$ the mobility rapidly decreases
with increasing $R_l$. 
In Fig.~\ref{fig:2}(b), there is a local minimum in
$P_{6}$ near $R_{l} = 1.0$ which 
coincides with the local maximum in the mobility 
shown in Fig.~\ref{fig:2}(a). 
For $R_{l} > 8.0$, $P_{6}$ begins to grow
rapidly with increasing $R_l$, indicating the onset of cluster formation.
The rise in $P_{6}$ corresponds with a decrease in the mobility
in Fig.~\ref{fig:2}(a). 
At very low $R_{l}$ the system 
acts like a zero temperature non-active system through which the 
probe particle can easily move.
As $R_l$ increases for fixed $\phi$, the probe particle encounters active
particles more frequently, decreasing its mobility.
The system becomes more disordered as $R_l$ increases, reaching a maximally
disordered state at $R_l=1$.  The liquid nature of the disordered state
permits the probe particle to move more easily through the sample, 
giving a small increase in the mobility.

Once clusters begin to form for $R_l>8$, the probe particle often becomes
trapped within a cluster.  During this trapped period, the motion of the 
particle in the driving direction drops nearly to zero, so that a probe
particle caught in a cluster is effectively in a locally jammed state.
Here, the term jamming refers to the impedance of motion due to steric
interparticle interactions.
Since the clusters are dynamical in nature, the driven particle
eventually escapes from the cluster and promptly moves much more rapidly
as it passes through the
low density liquid phase that surrounds 
the clusters. 
As $R_{l}$ increases, individual clusters become longer-lived, so the probe
spends larger amounts of time trapped inside clusters in a low mobility
state, resulting in a net decrease of the average mobility with increasing
$R_l$.

\begin{figure}
\includegraphics[width=3.5in]{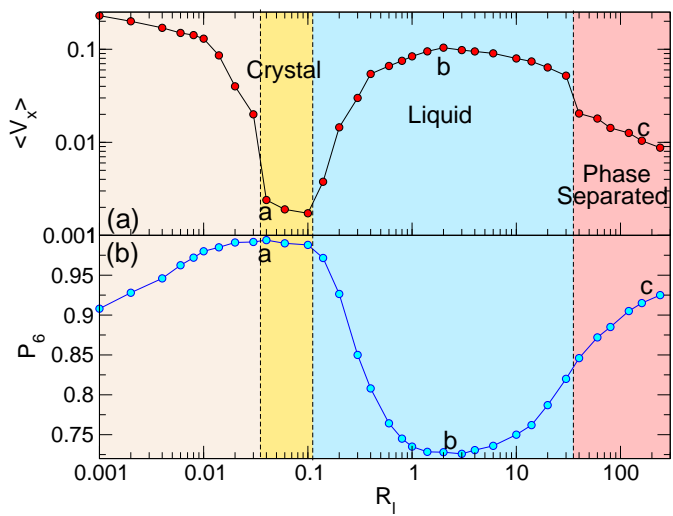}
\caption{ 
(a) 
$\langle V_x\rangle$ 
vs 
$R_{l}$ 
for a system with $\phi=0.801$.
(b) 
The corresponding 
$P_{6}$ vs $R_{l}$. 
Point {\bf a}, in the crystalline phase where the mobility is low,
is illustrated in Fig.~\ref{fig:4}(a).
Point {\bf b}, in the liquid phase where the mobility is high,
is illustrated in Fig.~\ref{fig:4}(b).
Point {\bf c}, in the clump or phase separated phase where the
mobility decreases again, is illustrated in Fig.~\ref{fig:4}(c).
}
\label{fig:3}
\end{figure}

The nonmontonic behavior of the mobility with $R_{l}$ also depends on $\phi$.  
In Fig.~\ref{fig:3}(a) we plot $\langle V_x\rangle$
versus $R_{l}$ for a system with $\phi = 0.801$. 
For $ 0.001 < R_{l} < 0.01$, there is 
a decrease in the mobility with 
increasing $R_{l}$ which becomes more pronounced for 
$R_{l} > 0.01$ before $\langle V_x\rangle$ reaches a local minimum
near $R_{l} = 0.005$. 
For $R_{l} > 1.0$ the mobility sharply increases 
with increasing $R_l$ to a local maximum near $R_{l} = 2.0$. 
The total drop and recovery of
$\langle V_x\rangle$ between $R_l=0.01$ and $R_l=1$ covers more than
an order of magnitude.
In Fig.~\ref{fig:3}(b) we plot the corresponding $P_{6}$ versus $R_{l}$, 
where for 
$0.001 \leq R_{l} < 0.01$, $P_{6}$ {\it increases} with increasing $R_{l}$,  
indicating that the activity is
producing additional order in the system. 
For $0.01 < R_{l} < 0.02$, there is a local maximum in 
$P_{6}$ with $P_6=0.99$, indicating 
almost perfect triangular ordering of the system. 
This same interval of $R_{l}$  also corresponds to the low mobility region
in Fig.~\ref{fig:3}(a), as highlighted by the vertical dashed lines. 
The peak in the mobility centered near $R_{l} = 2.0$ is matched by
a local minimum in $P_{6}$ in Fig.~\ref{fig:3}(b), 
indicating that the activity has now disordered the system. 
There is no sharp signature in $P_6$ at the onset of the phase separated
state, but $P_6$ increases with increasing $R_l$ in this regime as the
mobility drops.

\begin{figure}
\includegraphics[width=3.5in]{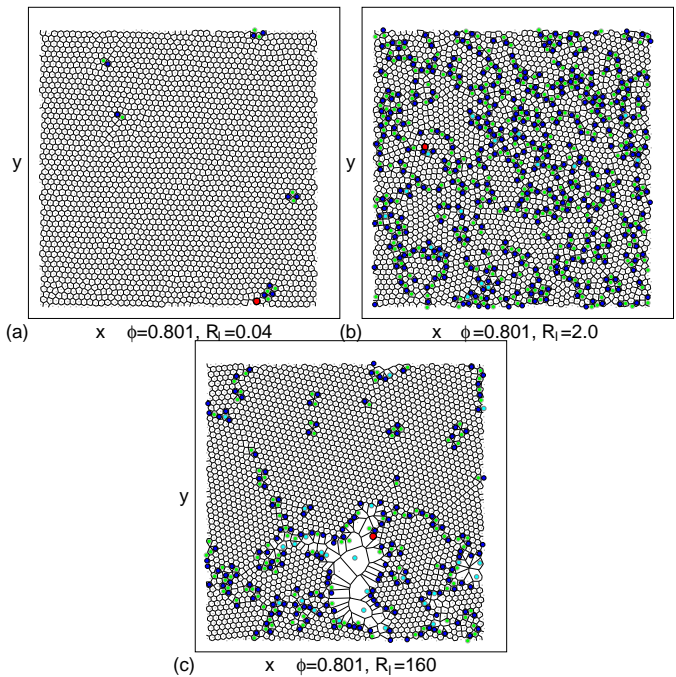}
\caption{Voronoi construction images
for the system in Fig.~\ref{fig:3} with $\phi = 0.801$. 
Particle coordination numbers: 
$z_i=5$ (dark blue), $z_i=6$ (white), $z_i=7$ (green), $z_i\geq 8$ (light blue).
The driven particle is marked with a red dot.
(a) At $R_{l} = 0.04$, labeled {\bf a} in Fig.~\ref{fig:3},
the system forms a crystallized state. 
(b) At $R_{l} = 2.0$, labeled {\bf b} in Fig.~\ref{fig:3}, 
the system forms a disordered uniform liquid state.
(c) At $R_{l} = 160$, labeled {\bf c} in Fig.~\ref{fig:3}, 
the system forms a phase separated state with high density 
clusters and a low density gas of particles.
}
\label{fig:4}
\end{figure}

Figure~\ref{fig:4}(a) 
shows the Voronoi construction obtained from the particle
positions for the system in Fig.~\ref{fig:3} 
at $\phi = 0.801$ and $R_{l} = 0.04$.
Although $\phi = 0.801$ 
is well below the non-active crystallization density of 
$\phi = 0.9$,  
we find
a uniform activity-stabilized triangular lattice 
interspersed with a small number of defects. 
The fact that the
probe particle mobility drops
nearly to zero in this regime indicates that 
this phase behaves like a solid. 
This 
crystalline state
is distinct from the cluster phase 
in that it is completely uniform and appears at much lower values 
of $R_{l}$. 
The crystallization occurs only when $\phi$ is relatively high, so that
each particle can be regarded, on average, as filling $\phi$ percent of
a hexagon of side $l_b$ centered on the particle:
$l_b=\sqrt{2 \pi r_d^2/(3 \sqrt{3}\phi)}$.  Then
the average surface-to-surface spacing between neighboring particles is
$d_s=l_b-r_d$.  For $\phi=0.801$, $d_s=0.11$.  Since all of the particles
are actively moving, a collision occurs every time the particles traverse
an average distance of $r_s=d_s/2=0.055$.  When $R_l \sim r_s$, there is
a matching effect in which each particle is struck, on average, nearly 
uniformly around its circumference by neighboring particles, resulting
in the formation of the crystalline state.
If $\phi$ decreases, $r_s$ becomes large enough that multiple collisions
between neighboring particles can occur as a particle moves the corresponding
distance $R_l$, so the crystalline ordering is lost.  
The crystallization is accompanied by a maximum in $P_6$ in
Fig.~\ref{fig:3}(b).
When
$R_l \ll r_s$, the system is in an
unjammed disordered state and the probe particle can 
move easily through the sample.

Although the activity can induce a crystallization when
$R_l \approx r_s$, if $R_l$ increases too much
the forces exerted on a given particle by its neighbors are
no longer symmetric; instead, an unbalanced excess force arises whenever a
long-lived contact with another particle forms for two particles swimming
in opposite directions toward each other.  Thus the crystalline order is 
lost when $R_l$ is too large, as illustrated in Fig.~\ref{fig:4}(b) for
$R_l=2.0$.
The disordering of the system appears as a drop in $P_6$ in Fig.~\ref{fig:3}(b),
and Fig.~\ref{fig:4}(b) shows that the sample enters a uniform 
active liquid state
containing numerous dislocations.
Since the system is now a liquid instead of a solid, the probe particle 
can pass through the sample much more easily and the mobility in 
Fig.~\ref{fig:3}(a) increases.
As $R_{l}$ is further increased, the system enters the cluster phase where 
large regions of the sample have sixfold ordering,
as illustrated in Fig.~\ref{fig:4}(c)  
for $\phi = 0.801$ and
$R_{l} = 160$. 
At the corresponding density in Fig.~\ref{fig:3}, $P_6$ is large again and the
mobility is low.
The decrease in average mobility occurs since the velocity of the probe
particle drops nearly to zero whenever it traverses a dense portion of
the sample.

\begin{figure}
\includegraphics[width=3.5in]{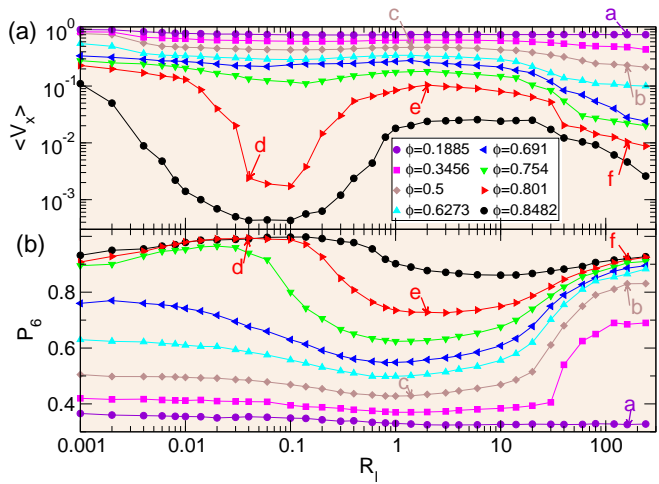}
\caption{ 
(a) Mobility $\langle V_x\rangle$ of the probe
vs run length $R_{l}$ for $\phi = 0.1885$, 0.3456, 0.5, 
0.6273, 0.691, 0.754, 0.801, and $0.8482$, from top to bottom. 
(b) 
Corresponding $P_{6}$ vs $R_{l}$ for
$\phi = 0.1885$, 0.3456, 0.5, 0.6273, 0.691, 0.754, 0.801, and $0.8482$, 
from bottom to top. 
}
\label{fig:5}
\end{figure}

In Fig.~\ref{fig:5}(a) we plot $\langle V_x\rangle$ versus $R_l$ for
$\phi$ ranging from 
$\phi = 0.1885$ to 
$\phi=0.8482$, 
while in Fig.~\ref{fig:5}(b) we show corresponding $P_6$ versus $R_l$.
There is a pronounced minimum in $\langle V_x\rangle$
for the $\phi = 0.801$ and $\phi=0.8482$ curves due to the
occurrence of activity-induced crystallization as described above.
A small residue of this effect 
remains at $\phi = 0.754$. 
For $1 < R_l < 30$, the system is in a disordered liquid state
at all the values of $\phi$, 
while for $R_{l} > 30$, the increase in
$P_{6}$ indicates that the system is entering the clump phase in which
the mobility drops. 
For $\phi = 0.1885$, the system remains in the liquid phase 
over the entire range of $R_{l}$ and shows little change in $\langle V_x\rangle$
or $P_{6}$. 
These results indicate that the activity induced crystallization 
at low $R_l$ 
only occurs when $\phi$ is sufficiently large,
whereas the cluster phase at large $R_l$ appears over a
wide range of $\phi$. 

\begin{figure}
\includegraphics[width=3.5in]{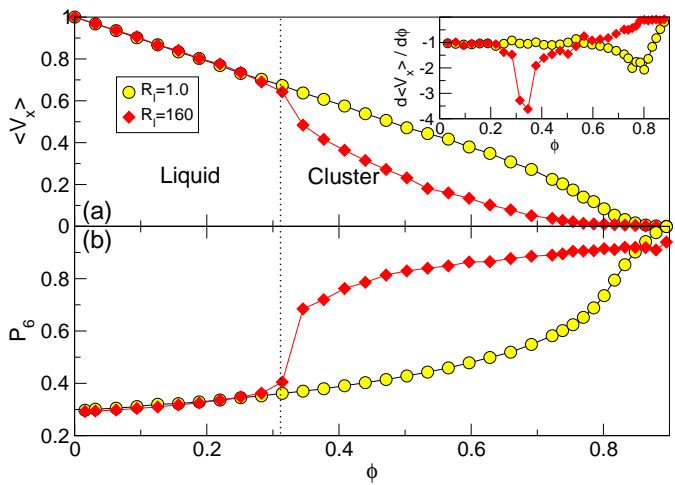}
\caption{
(a) Mobility $\langle V_x\rangle$ vs $\phi$ for $R_l=1.0$ (circles) and
$R_l=160$ (diamonds).
Inset: corresponding $d\langle V_x\rangle/d\phi$.
(b) $P_{6}$ vs $\phi$. 
Dashed line indicates the correspondence between the
clustering onset and the mobility drop.
}
\label{fig:6}
\end{figure}

To further characterize the change in the mobility with the onset of 
clustering and ordering, 
in Fig.~\ref{fig:6} we plot $\langle V_x\rangle$
and $P_6$
versus $\phi$ for samples with $R_{l} = 160$ and
$R_{l} = 1.0$.
For $0 < \phi < 0.3$, $\langle V_x\rangle$ is
independent of $R_{l}$ and decreases linearly 
with increasing $\phi$. 
For $\phi > 0.3$, the $R_l=160$ system undergoes a 
transition to the cluster state, 
producing a sharp increase in $P_{6}$ 
and a simultaneous mobility drop.
As $\phi$ increases, $\langle V_x\rangle$ in the $R_{l} = 1.0$ 
system decreases linearly,
while in the $R_l=160$ system $\langle V_x\rangle$
drops rapidly in the cluster phase until becoming nearly zero
for $\phi >  0.75$. 
In the inset of Fig.~\ref{fig:6}(a),
the plot of $d\langle V_x\rangle/d\phi$ versus $\phi$ shows
the much earlier mobility drop of the $R_l=160$ system compared
to the $R_l=1.0$ system, where the drop does not occur until 
$\phi \approx 0.8$. 
Near $\phi=0.6$, Fig.~\ref{fig:6}(a) shows that
$\langle V_x\rangle$ in the $R_l=160$ system is more than ten times
smaller than in the $R_l=1$ system.
Close to the hard sphere crystallization
density  of $\phi=0.9$, the two systems have nearly equal mobilities.

\begin{figure}
\includegraphics[width=3.5in]{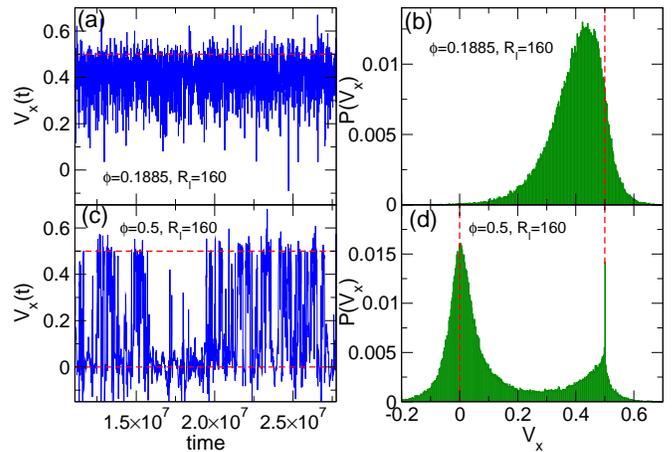}
\caption{
(a) Portion of a probe velocity time series 
$V_{x}(t)$ in the uniform liquid state at $\phi=0.1885$ and $R_l=160$. 
The dashed line indicates the free probe velocity $V_0=F_d=0.5$.
(b) The probability distribution of 
the velocity fluctuations $P(V_{x})$ from (a). 
(c) Portion of $V_x(t)$ in the phase separated state 
at $\phi=0.5$ and $R_l=160$
where
two-level fluctuations occur. 
(d) $P(V_{x})$ from (c) 
showing peaks at $V_{x} = 0.0$ and $V_x=V_0=0.5$.   
}
\label{fig:7}
\end{figure}

\section{Velocity Fluctuation Distributions}
We next examine the velocity fluctuation distributions of 
the probe particle. 
In 
Fig.~\ref{fig:7}(a) we plot
a representative portion of
the probe velocity time series $V_{x}(t)$ in the
liquid phase from Fig.~\ref{fig:1}(a) at $\phi = 0.1885$ and $R_{l} = 160$. 
The dashed line in Fig.~\ref{fig:7}(a) indicates the velocity 
$V_0=F_d=0.5$ at which
the probe particle would move in the absence of the active bath particles. 
Figure~\ref{fig:7}(b) 
shows that the corresponding probability distribution function $P(V_{x})$ 
has a skewed Gaussian shape 
with a maximum near $V_{x} = 0.425$.  
We find similar $P(V_x)$ distributions 
in other uniform liquid states. 
For lower values of $\phi$, a peak in $P(V_x)$ at $V=V_0$ begins to
emerge when
the probe particle moves large distances before encountering 
another particle. 
Figure~\ref{fig:7}(c) shows $V_x(t)$
for $\phi  = 0.5$ and $R_{l} = 160$ when the sample is in
the cluster phase illustrated in Fig.~\ref{fig:1}(b).
Here the noise fluctuations are of two-level or 
telegraph type, and
$V_{x}$ jumps between the values
$V_{x} = 0.0$  and $V_{x} = V_0=0.5$, highlighted by the dashed lines. 
The resulting $P(V_x)$ contains two clear peaks, as shown in 
Fig.~\ref{fig:7}(d).
The two-level behavior arises because the probe velocity is nearly zero when
the probe becomes trapped in a cluster, and nearly $V_0$ when
the probe moves through the low density gas surrounding the clusters
where it undergoes very few collisions. 
The peak at $V_{x} = 0.5$ for $\phi=0.5$ in 
Fig.~\ref{fig:7}(d) is quite sharp, whereas the
peak in $P(V_{x})$ at $\phi = 0.1885$ 
in Fig.~\ref{fig:7}(b) is centered at 
$V_x=0.425$ even though the system is in a low density 
liquid phase. 
The difference in peak sharpness arises because the liquid density 
at $\phi=0.1885$, although low, is considerably higher than the
density in the gas state surrounding the clusters that form
at $\phi=0.5$, and the resulting larger number of probe-bath particle encounters
in the liquid state broaden the peak relative to its width in the
gas state.
We observe similar
telegraph noise distributions in the other phase separated regimes for 
$R_{l} = 160$ over the range 
$0.377 < \phi < 0.75$, and find that the weight of the distribution shifts from 
the $V_x = 0.5$ peak to the
$V_x = 0.0$ peak as $\phi$ increases.  
The telegraph noise signal can 
be viewed as a linear combination
of the velocity fluctuation distributions 
from the low density phase,
which has a skewed Gaussian shape with a peak near the driving value 
of $F_{d} = 0.5$,
and the high density phase, which has
one peak centered at zero and a second peak centered at $0.5$, 
similar to the shape shown in
Fig.~\ref{fig:7}(d).

\begin{figure}
\includegraphics[width=3.5in]{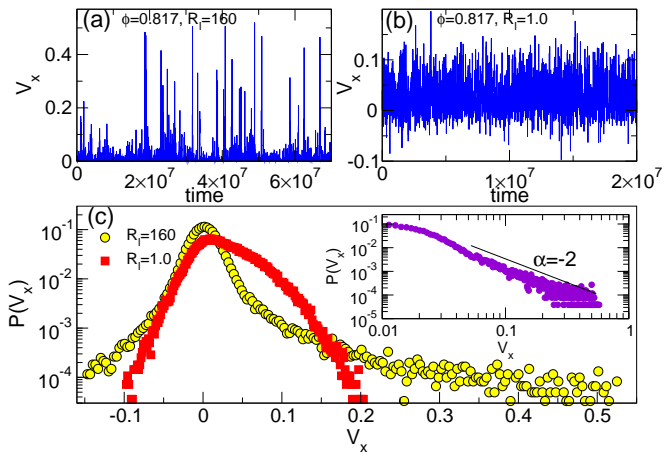}
\caption{
(a) Portion of $V_{x}(t)$ in the active jamming phase
at $\phi=0.817$ and $R_l=160$
showing that the probe particle moves in discrete jumps
or avalanches. 
(b) $V_x(t)$ 
in the uniform liquid state at $\phi=0.817$ and $R_l=1.0$
where the probe
moves continuously.
(c) Log-linear plot of $P(V_{x})$ from the time series at 
$\phi = 0.817$ with $R_l=160$ (circles) and $R_l=1.0$ (squares). 
Inset: log-log plot of $P(V_x)$ for the positive $V_x$ values 
at $R_{l}  = 160$. Solid line: a power law fit with 
exponent $\alpha = -2.0$.  
}
\label{fig:8}
\end{figure}

For $\phi > 0.75$ and $R_{l} = 160$, the probe is mostly stationary 
and  only 
moves in short bursts or avalanches with a broad distribution of jump sizes.
This is illustrated 
by the plot of
$V_{x}(t)$ in Fig.~\ref{fig:8}(a) for $\phi = 0.817$ and $R_l=160$.
If $\phi$ is held fixed and $R_l$ is reduced to $R_l=1.0$,
the system 
forms a disordered liquid state and the 
probe motion is no longer intermittent, as
shown in the plot of $V_x(t)$ in Fig.~\ref{fig:8}(b). 
Figure~\ref{fig:8}(c) illustrates the corresponding 
$P(V_{x})$ curves on a log-linear scale for $R_{l} = 1.0$ and $R_{l} = 160$.
For $R_{l} = 160$, $P(V_x)$ has a pronounced peak near $V_{x} = 0.0$
and a broad 
tail for $V_x>0.1$,
while for $R_{l} = 1.0$, 
$P(V_x)$ falls off more rapidly 
for $V_x>0.1$ and the 
maximum is centered slightly higher than $V_{x}  = 0.0$.  
The inset of Fig.~\ref{fig:8}(c) shows a log-log 
plot of $P(V_x)$ for positive values of $V_{x}$ 
at $R_{l} = 160$.  The solid line is a power law fit to 
the form $P(V_x) \propto V_x^{-2}$.  Similar fits 
can be made for $\phi > 0.75$ and $R_{l} = 160$.  

We characterize
the $\phi > 0.75$ and $R_{l} > 30$ regime as actively jammed, since
this
is where the probe particle velocity is power-law distributed.
For non-active matter systems very near the jamming transition, a driven
probe particle also moves in an intermittent fashion and undergoes
avalanches with a size distribution that can be fit to a power law  
\cite{17,18}. 
The active matter avalanche behavior suggests that the 
active system shares characteristics with a jammed phase, but that these
characteristics arise at densities well below the non-active jamming
density.
Jamming behavior of non-active disks has been widely studied for 
bidisperse, rather than monodisperse, disk sizes, which produce
a disordered structure at high densities.  Jamming has been used as a
general concept for describing amorphous solids.
In the phase-separated active matter system, the
dense regions of the sample are polycrystalline rather than amorphous;
however, the polycrystalline
structures are not 
static but dynamically change, 
so that the system can be regarded as dynamically amorphous. 
Images of the states that form at high $\phi$ and large $R_l$ show the
formation of
various grain boundary structures, 
implying that a probe motion avalanche might occur whenever a grain
boundary passes over the probe particle.
There are, however, also other types of defects
and small voids present that can also affect the probe particle motion.  
In addition, 
the probe particle can itself nucleate local topological defects and
voids, which then interact with the motion of the probe particle.
It is beyond the scope of this work to definitively determine
whether the dense active phase is truly in a critical
state; however, the observed power-law
exponent of $\alpha=-2.0$ is consistent with the
class of time-directed 
avalanche systems \cite{39}. 
Future  directions include studying
polydisperse active particle assemblies, active rods, or active dumbbells.

\section{Conclusion} 

In conclusion,
we have examined the mobility and velocity fluctuations of an 
externally driven probe particle moving through
a bath of active matter disks for varied activity and particle 
density.
As a function of increasing run length, we show that 
there is a pronounced  drop in the probe particle mobility 
at the transition from a uniform liquid state to a 
cluster or phase separated state. 
The mobility reduction
in the cluster state arises due to the temporary trapping of the probe
particle by dense clusters.
When the probe particle escapes a cluster, it moves with a much 
higher, nearly free
mobility
until it encounters another cluster. 
The time series
of the velocity in the uniform liquid regime exhibits 
a skewed Gaussian shape, while in the phase separated regime
we observe
two-level velocity fluctuations 
as the particle jumps between 
dense low mobility regions of the sample and gas-like high mobility regions
of the sample.
As the density of active particles increases, we find
strongly nonmonotonic behavior of the probe particle mobility as a function of 
run length.  
For finite but small run lengths, the mobility initially drops by nearly
two orders of magnitude when 
the system enters an activity-induced crystallization regime. 
As the run length increases, 
the crystalline state becomes unstable, the system disorders, and the 
mobility can increase by a few orders of  magnitude. 
At the largest run lengths the system enters the phase separated regime 
and the mobility of the probe particle decreases again. 
We also find that in samples with high densities and large run lengths,
the probe particle moves in an
intermittent fashion 
via discrete jumps or avalanches, and that the probe velocity
is power-law distributed. 
The avalanche velocity distributions suggest that the dense active 
system may exhibit 
a critical behavior similar to that found for a probe particle 
moving through a non-active disordered 2D assembly
just below the  
jamming transition.
Our results should be general to both run-and-tumble active systems as well 
as active Brownian particles.     

\acknowledgments
This work was carried out under the auspices of the 
NNSA of the 
U.S. DoE
at 
LANL
under Contract No.
DE-AC52-06NA25396.

\end{document}